\documentclass{ceurart}

\begin{document}

\copyrightyear{2026}

\copyrightclause{Copyright for this paper by its authors. Use permitted under Creative Commons License Attribution 4.0 International (CC BY 4.0).}

\conference{Preprint. Submitted to HUMANE-LLM 2027; under review. Not yet accepted or published.}

\title{ARRANGE: Agentic Skills for Layering Vocals}

\author[1]{Henry Tan}[email=henrytan@g.harvard.edu]

\author[2]{Eidee Tan}[email=eidee@enrica.ai]

\address[1]{Harvard University, United States}

\address[2]{enrica.ai}

\begin{abstract}
Adding a singing voice to an existing piano recording requires decisions about melody, syllable placement, pronunciation, timing, and vocal register. ARRANGE organizes this work as a sequence of agent skills with readable intermediate artifacts and composer approval checkpoints. An LLM translates instructions in plain English, notes played on a keyboard, and vocal references into local edits; deterministic scripts validate and apply those edits before open audio models render the result. The Score Canvas makes note, syllable, and pronunciation decisions inspectable and editable. In a benchmark of 16 revision requests, three draws per request, and five LLMs, typed edits achieved exact-change correctness of 0.83–0.98, compared with 0.33–0.69 for whole-plan rewriting without ARRANGE instructions. A separate score-editing benchmark and a first-person worked example examine the interaction between representations and corrections. Author listening evaluations further explore voice transfer and register selection, including 185 listening presentations across eight copied voices. These evaluations support the value of constrained editing and expose limits of both model execution and listening measures. The contribution is an interaction architecture for retaining musical decisions through inspectable intermediate representations, rather than a new singing model or evidence of general usability.
\end{abstract}


\maketitle

\hypersetup{pdfsubject={Preprint submitted to HUMANE-LLM 2027; under review; not yet accepted or published},pdfauthor={Henry Tan; Eidee Tan},pdftitle={ARRANGE: Agentic Skills for Layering Vocals},pdfkeywords={human-AI co-creation, agent skills, mixed-initiative interaction, user agency, singing voice synthesis, music notation}}

\section{Introduction}

A pianist records a piece and writes words for it. They want those words sung over their own playing. This gives them a rapid feedback loop. Producing that result involves several small decisions: which notes the voice sings, which syllables fall on those notes, where each line enters against the performance’s timing, and how each syllable is pronounced. A convincing audio result does not necessarily make these decisions easy to inspect or revise.

The author first explored this task with Suno. In that trial, the generated arrangement replaced the recorded piano, and revising individual musical decisions through prompting became difficult. This experience motivates ARRANGE, but it does not establish a general limitation of current commercial tools. Suno now documents uploaded audio, multitrack editing, clip pitch controls, and time-stretch settings \cite{ref12}. Open systems also support editing and audio conditioning \cite{ref2,ref3,ref14,ref15}. Our question is therefore more specific: how can an LLM-based interface expose and preserve the composer’s decisions while adding a voice to a recorded performance?

ARRANGE treats the task as compilation through agent skills. The composer supplies plain English, a melody played on a keyboard over the recording, and vocal references. Each stage produces an intermediate artifact that can be inspected and patched before work continues. When a syllable is missing or a note is wrong, the target of correction is the artifact that represents that decision. The agent translates the request; scripts constrain its application; the composer judges the musical result.

The compiler analogy concerns explicit intermediate representations, validation, and local transformations. Natural-language requests remain ambiguous, and singing models remain stochastic. ARRANGE does not make either deterministic. It moves decisions that can be represented explicitly into a shared score and typed edit vocabulary, while leaving approval with the composer.

The contributions are:

\begin{enumerate}

\item A six-stage workflow that assigns each skill an input, output, and composer checkpoint, with audio backends separated from the interaction contract.

\item The Score Canvas, which connects notation, syllables, pronunciation, and playback to support local correction.

\item An evaluation combining plan-revision and score-editing benchmarks, a first-person worked example, and author listening evaluations that reveal the strengths and limitations of the approach.

\end{enumerate}

The intended user in this work is a composer who can read notation and play a melody on keys. We evaluate that setting rather than claiming accessibility across musical backgrounds.

\section{Related work}

\textbf{Visible LLM workflows.} AI Chains demonstrated how editable intermediate results can improve transparency and controllability in multi-step LLM interaction \cite{ref13}. Agent Skills package instructions, scripts, and resources that an agent can load for a task \cite{ref1}. ARRANGE uses this packaging and specializes its intermediate representations for musical decisions. Its typed operations act on artifact state rather than asking the model to reproduce the entire artifact. Work on atomic concept edits concerns controlled changes to model inputs \cite{ref5}; ARRANGE instead studies the scope of changes applied to a composer’s plan or score.

\textbf{Music generation and editing.} Suno generates songs from text and supports audio-based production and editing features \cite{ref12}. YuE addresses long-form lyrics-to-song generation \cite{ref14}; ACE-Step and ACE-Step 1.5 support generation and editing operations \cite{ref2,ref3}. YuE2 combines symbolic and audio generation \cite{ref15}, so symbolic planning before synthesis is not unique to ARRANGE. Our emphasis is the interaction contract: a composer-readable artifact, explicit correction operations, and an approval point between stages.

\textbf{Audio and score backends.} High-resolution piano transcription provides notes and timing from a recorded performance \cite{ref6}. SoulX-Singer supports score-conditioned singing synthesis \cite{ref9}, while Seed-VC supports voice conversion, including singing with pitch conditioning \cite{ref7}. Optical music recognition offers a possible additional entry point from an existing score \cite{ref10}. These are complementary components; ARRANGE does not contribute a new transcription, separation, synthesis, or conversion model.

\textbf{Human–AI co-creation.} AI Song Contest teams described dividing creative work into parts and negotiating the limitations of models \cite{ref4}. Studies of text-to-music production identify difficulties integrating generated material into musicians’ workflows \cite{ref11}. Detailed accounts of co-creative songwriting also make human contributions visible \cite{ref8}. ARRANGE builds this decomposition into an agent workflow and uses notation as the shared representation through which the composer can contest and correct decisions.

\section{Skills as compilation stages}

An ARRANGE skill is a folder of instructions, scripts, and references. Six conceptual stages organize the workflow; some share an implementation folder. Table 1 separates each stage’s artifact from the decision the composer makes about it.

\begin{table}[ht]

\caption{Stage contracts in ARRANGE. The checkpoints are instructions governing the workflow. They should not be interpreted as a formally verified guarantee that an unconstrained agent cannot bypass them.}

\centering

\small\setlength{\tabcolsep}{3pt}

\begin{tabular}{@{}p{.12\linewidth}p{.20\linewidth}p{.30\linewidth}p{.29\linewidth}@{}}

\toprule

\textbf{Stage} & \textbf{Input} & \textbf{Output} & \textbf{Composer checkpoint} \\

\midrule

Listen & Piano recording and tapped beats & Transcribed notes, key estimate, and performance-relative beat map & Do the bar numbers match how I count the piece? \\

Play the Line & Recording and beat map & Melody take containing pitch and rhythm played on keys & Is this the intended line, or should a phrase be replayed? \\

Fit & Melody take and words & Notes with syllables and per-note pronunciation & Is every intended syllable present, on the right note, with the right sounds? \\

Plan & Approved musical material and language requests & Layer, entry, and take-selection plan & Does the proposed change match the request and retain earlier decisions? \\

Sing & Approved score and voice reference & Vocal render aligned to the recording’s clock, with entry measurements & Does it sing the intended line in a suitable register? \\

Mix & Recording and approved vocal render & Mix, stems, and DAW session & Are balance and placement acceptable? \\

\bottomrule

\end{tabular}

\end{table}

The agent selects skills, translates language, and calls scripts and audio models. We call its structured revision commands typed edits: each selects a predefined operation and supplies named arguments with specified types. For example, “bring the strings in two bars later” becomes the typed edit \texttt{shift\_entry(layer="strings", by\_bars=2)}. Here, \texttt{shift\_entry} is the operation, \texttt{layer} identifies the target layer, and \texttt{by\_bars} specifies the shift in bars. The model outputs this command; an application script checks the operation and arguments and applies it to the existing plan. The model therefore specifies a local change rather than rewriting the whole plan. Unsupported or ambiguous requests are intended to produce a question rather than an inferred arrangement decision.

This separation limits the amount of state a model must regenerate. Validation checks whether a typed edit follows the operation’s syntax and argument rules; it does not establish that the model interpreted the musical request correctly. For example, a valid \texttt{shift\_entry} command could target the wrong layer or specify the wrong shift. The revision benchmark therefore measures both intended changes and collateral changes. It also records modifications to composer-owned fields that the expected edit did not touch.

The Score Canvas and the plan are related intermediates rather than one interchangeable file. The lyric skill owns the performed melody and note-level fit; the plan skill schedules layers and records arrangement choices. Their evaluation tasks are reported separately. Listen uses piano transcription \cite{ref6}; Sing uses SoulX-Singer \cite{ref9} or a voice-conversion route such as Seed-VC \cite{ref7}. Backend replacement requires an adapter that respects the score and timing interface. It does not imply that arbitrary models are interchangeable without engineering work.

The accompaniment remains the source recording rather than a newly generated performance. A non-octave change of key can require pitch-shifting that recording; retaining the performance therefore does not mean that the final waveform is always unchanged.

\section{The Score Canvas}

\begin{figure}[ht]

\centering

\includegraphics[width=\linewidth]{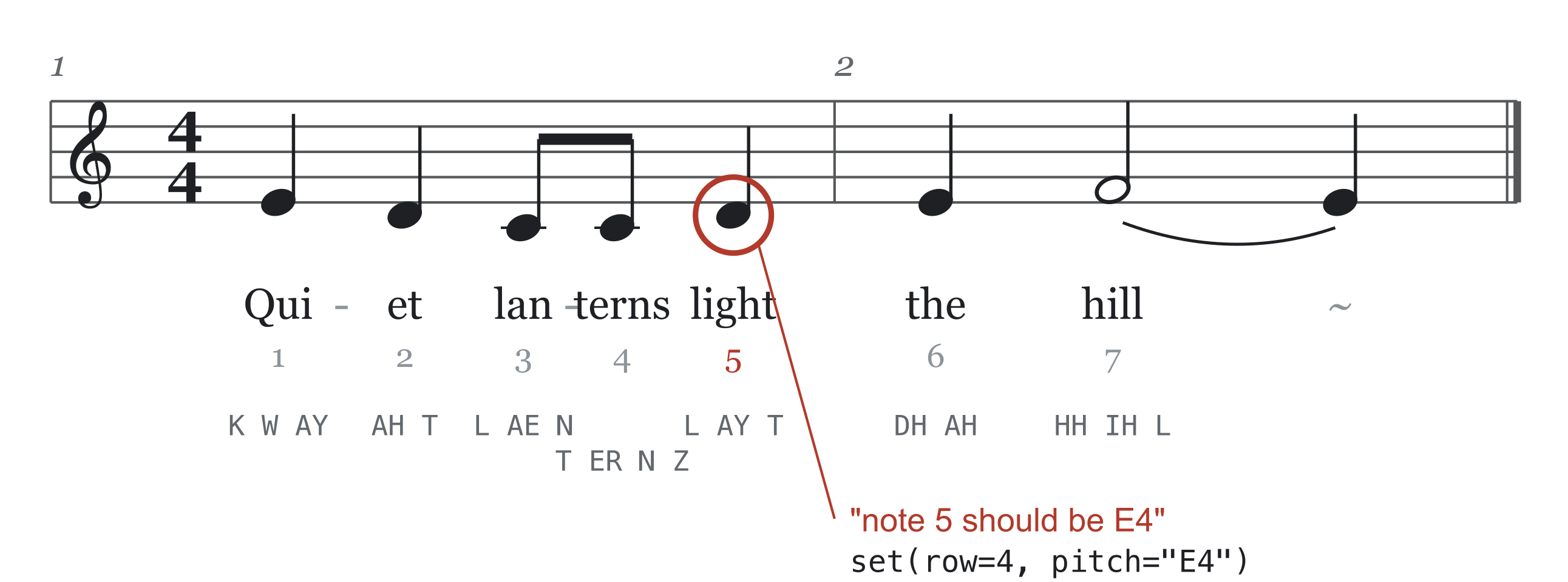}

\caption{Schematic score example from the ARRANGE project page. Notes share syllable labels, display indices, and per-note sounds; the marked note is the target of a local pitch request. This illustration is not a screenshot of the evaluated 15-bar excerpt.}

\label{fig:canvas}

\end{figure}

The Score Canvas presents the fit as engraved notation with words under notes. Bar numbers follow the composer’s tapped beats, each syllable has a visible index, and a pronunciation view shows the sounds supplied to the singer. Playback from a note connects the symbolic representation to the piano and voice. These references let the composer and agent point to the same musical event.

\subsection{Playing and identifying the line}

The composer plays the intended vocal melody on keys while hearing the recording. This captures pitch and rhythm together, rather than requiring the line to be entered as text or corrected from an automatically transcribed singer. In the author’s use, this was preferable to dragging syllables on a timeline, correcting a reference transcription, or typing the melody. This is a first-person design observation, not a comparative usability result.

A request can identify its target by bar and beat, syllable index, word, pitch, or an image of the score. The corresponding edit vocabulary includes setting a field, moving or shifting a note, inserting or deleting a row, changing pronunciation, asking a question, and answering a question. A short display index can be translated into an internal row index; the display and file representations must remain consistent.

\subsection{Locating and correcting errors}

Three faults in the worked example illustrate why the intermediate matters. First, splitting a consonant onto its own note produced an unwanted vowel-like sound in “morning.” Second, a line with more syllables than available notes left words unsung. Third, the split of “Footsteps” across repeated notes placed stops at the join and made the word sound broken. Each fault was identified at the note or pronunciation level and corrected locally.

The pronunciation view is consequential because a text change alone may not change the sounds sent to a synthesizer. For “Footsteps,” the composer adjusted the sound split across notes so that one stop consonant carried into the second note. The fit and pronunciation file retained that choice for subsequent renders.

These examples motivate visibility at the level of a failure, rather than only a summary of what an agent intended. They do not establish that every audible defect can be diagnosed from notation. A singing model may still render a correct score poorly, so each changed line must be heard again.

\section{Evaluation}

We ask four questions: whether typed operations preserve intended plan changes; whether score-oriented representations help execute corrections; what the workflow affords in one composer’s practice; and what author listening evaluations reveal about voice transfer and register selection. Model-output benchmarks are not user studies. The author is also the composer and listener throughout the musical evaluations.

\subsection{Plan revision across five models}

The benchmark contains 16 frozen revision requests: seven on the author’s real arrangement plan and nine on a synthetic plan. Each request has an expected change. Each model answered it three times in four conditions, giving 48 draws per model and condition, or 960 draws in the five-model comparison.

The conditions were: typed ARRANGE operations applied by a script; path-and-value patches; whole-plan rewriting with ARRANGE’s plan documentation; and whole-plan rewriting with a one-line instruction and no ARRANGE documentation. The first three conditions received the plan structure information. The comparison therefore evaluates an instruction, representation, and execution package; it does not isolate a vocabulary effect from all validation behavior.

Correctness requires exactly the expected changed paths, or a question without editing when the request is ambiguous or unsupported. Separate measures count unrelated changed paths, changes to protected composer-owned fields, invalid outputs, and missed re-render dependencies. Derived timing and ownership bookkeeping are excluded from the change counts.

\begin{table}[ht]

\caption{Exact-change correctness in the five-model plan benchmark. Each rate is based on 16 requests × 3 draws. The final column compares typed operations with rewriting without ARRANGE documentation. Wilcoxon tests pair the mean correctness of each request; p-values are exploratory and not adjusted for multiple comparisons.}

\centering

\small\setlength{\tabcolsep}{3pt}

\begin{tabular}{@{}p{.22\linewidth}p{.10\linewidth}p{.10\linewidth}p{.13\linewidth}p{.13\linewidth}p{.18\linewidth}@{}}

\toprule

\textbf{Model} & \textbf{Typed operations} & \textbf{Path patch} & \textbf{Rewrite with documentation} & \textbf{Rewrite without documentation} & \textbf{Difference and p} \\

\midrule

Claude Sonnet 5.5 & 0.98 & 0.81 & 0.38 & 0.33 & +0.65; p = 0.001 \\

Claude Opus 5.5 & 0.90 & 0.73 & 0.71 & 0.69 & +0.21; p = 0.031 \\

Claude Haiku 5.5 & 0.83 & 0.75 & 0.73 & 0.62 & +0.21; p = 0.047 \\

Gemma 4 31B & 0.90 & 0.67 & 0.62 & 0.46 & +0.44; p = 0.009 \\

Qwen 3.5 27B & 0.83 & 0.69 & 0.71 & 0.50 & +0.33; p = 0.078 \\

\bottomrule

\end{tabular}

\end{table}

Typed operations had the highest correctness for every model. For Sonnet, rewriting with or without documentation changed means of 19.9 and 22.1 unrelated fields per draw, respectively; the documented rewrite condition produced 11 invalid outputs out of 48. Across the five typed conditions, two protected-field changes were recorded, both for Qwen. Haiku’s typed condition still changed unrelated fields, showing that typed operations do not eliminate semantic mistakes.

An earlier three-condition Sonnet run achieved 0.96 correctness with typed operations and no collateral changes; the later four-condition run in Table 2 achieved 0.98. These are separate runs, not two estimates of the same set of outputs.

A follow-up tested whether a frontier model made ARRANGE’s task structure redundant. Claude Fable 5.1 achieved exact-change correctness of 0.94 with ARRANGE’s typed operations versus 0.69 with rewriting without ARRANGE instructions (48 draws per condition; paired Wilcoxon p = 0.062). The observed gap was smaller than Sonnet’s, but frontier capability alone did not match performance with ARRANGE on these requests.

Chained sessions tested eight successive requests on one real and one synthetic fixture, with three repetitions of each. Typed conditions ended exactly on the expected plan in three of six sessions per model. Excluding one beat-map request whose required timing was absent from the plan, all six sessions per model matched the expected state. This exclusion reveals an input limitation and should not be read as success on the complete session task.

Claude models were called through the Claude Code CLI with tools disabled for the single-completion benchmark. The open models were served locally through vLLM using FP8 checkpoints and server-default sampling, with a 32,000-token completion limit. The repository identifies the checkpoints and invocation commands; exhausted Qwen reasoning calls were scored invalid.

\subsection{Executing corrections on the score}

A replay study used 15 scored corrections from the composer’s own interactions. Most described an audible symptom rather than a fully specified edit. With the lyric skill’s structured representation, the three Claude models made an accepted fix in 0.42–0.47 of draws, or 0.53–0.69 when justified questions were included, compared with 0.18–0.31 from raw rows. The paired comparison of typed and raw-row conditions gave p = 0.064. This is suggestive evidence rather than a statistically established improvement. Pronunciation corrections succeeded in 26 of 36 typed draws and none of 36 raw-row draws; the structured condition exposed pronunciation information absent from the raw condition.

The public release contains nine of these correction requests. On that subset, accepted-fix rates are 0.56–0.59 for typed corrections and 0.26–0.41 for raw rows. The full-study and released-subset results should not be interchanged.

A separate curated benchmark contains 62 requests across three songs, including nine intentionally incomplete requests. Requests identify notes through score references or screenshots and cover local edits, patterns, passages, symptoms, revisions, and questions. Each model received three draws per request. Success requires the expected pitch and text, timing within specified tolerances, or the appropriate clarification or answer.

\begin{table}[ht]

\caption{Correctness on curated score edits. “Typed prompt” supplies the structured instructions in one prompt. “Installed skill” lets a Claude Code agent load the skill and use scripts; it is a separate execution condition. Each model-condition contains 186 draws.}

\centering

\small\setlength{\tabcolsep}{3pt}

\begin{tabular}{@{}p{.36\linewidth}p{.18\linewidth}p{.23\linewidth}p{.17\linewidth}@{}}

\toprule

\textbf{Model} & \textbf{Typed prompt} & \textbf{Raw rows and rewrite} & \textbf{Installed skill} \\

\midrule

Claude Haiku 5.5 & 0.80 & 0.60 & 0.84 \\

Claude Opus 5.5 & 0.92 & 0.83 & 0.89 \\

Claude Sonnet 5.5 & 0.91 & 0.74 & 0.82 \\

Gemma 4 31B & 0.55 & 0.49 & — \\

Qwen 3.5 27B & 0.68 & 0.58 & — \\

\bottomrule

\end{tabular}

\end{table}

Typed prompts improved the reported rate for every model. Request-level sign tests gave p < 0.001 for Haiku and p = 0.004 for Sonnet. The installed-skill results show that a constrained single-response benchmark is not a substitute for evaluating agent execution: loading instructions and using tools can introduce additional failures.

Fable also scored 0.90 with typed prompts versus 0.79 from raw rows on these 62 requests (186 draws per condition; request-level sign-test p = 0.115); its installed-skill score was 0.88.

\subsection{A worked verse on the author’s recording}

The worked example tested whether a composer could hear a vocal line over his own accompaniment and correct it through the score. It used 15 bars of the author’s piano recording of the theme from \emph{The Color Purple} by Quincy Jones, Rod Temperton, and Jeremy Lubbock, with words written by the author. He played the intended vocal melody on keys and tapped 88 beats to align it with the performance.

Listening exposed the three lyric and pronunciation faults described in Section 4. The composer located these in the score, made local corrections, and listened again after a re-render taking approximately one minute. Across five renders, line entries were within 0.1 s of their score positions. These timing and latency observations apply to one verse and setup. The next evaluation examines whether the chosen voice and key make the specified line convincing.

\subsection{Matching sampled voices to songs}

The usefulness of a sampled voice depended on its match to the song and key. In the author’s trials, some renders sounded thin or strained when the melody sat too high, and the voice became less recognizable. The listening experiments therefore examined which open audio backend produced useful vocals and how choosing a suitable key could improve the same sampled voice.

The author first compared SoulX-Singer and Seed-VC, judging both resemblance to the sampled voice and singing quality. Across 35 blind-coded clip presentations, the reported mean ratings were 95.9/100 for SoulX-Singer and 88.5/100 for Seed-VC. These are one listener’s judgments, without calibration across listeners. The author also observed that vocal color transferred more convincingly than singer-specific techniques such as belting, runs, and vibrato in the tested configurations.

Choosing a suitable key made some renders fuller, more recognizable, and more acceptable to the author. To investigate a useful starting point for this choice, eight copied voices sang key variants of the same 15-second excerpt of \emph{All the Things You Are}. The author heard 185 presentations across a pilot, shuffled blind listening, rechecks, and ordered comparisons. His final side-by-side selections served as the reference keys for comparing suggestions.

The tested rule matched the melody’s highest note to the upper part of the reference voice’s observed pitch range, represented by its 90th-percentile sung pitch. This percentile was selected in the pilot and fixed before the blind round. The rule’s suggestions differed from the author’s choices by 1.50 semitones on average across eight voices, compared with 6.00 for matching median pitches (Figure 2); individual errors reached four semitones. On the six voices used in the matched LLM comparison, the rule’s mean error was 1.33 semitones, while model-condition means were 4.0–7.4 semitones too high.

The listening context also mattered. For 52 identical audio files heard individually and in ordered key sequences, average strain ratings were about seven points lower in the latter setting. Numerical ratings therefore need consistent listening conditions to be interpretable. The rule proposes a key to audition; the composer still judges how the voice works with the song.

\begin{figure}[ht]

\centering

\includegraphics[width=\linewidth]{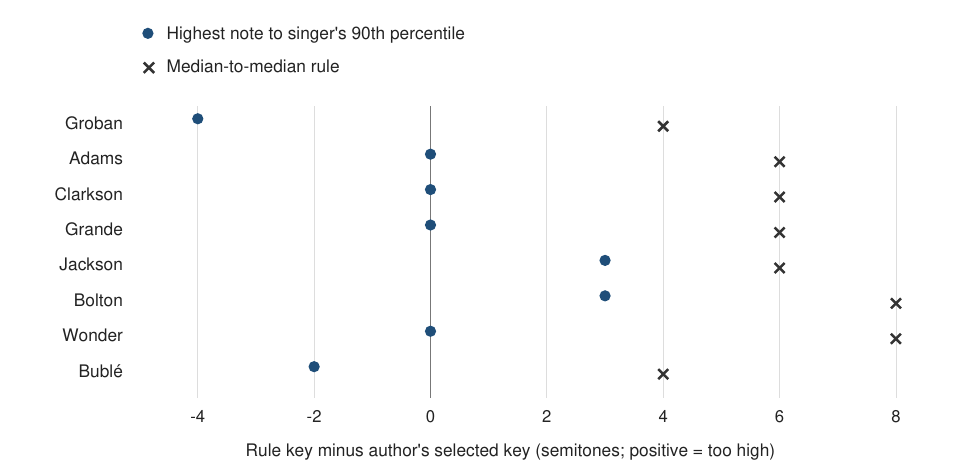}

\caption{Register-rule errors relative to the author's final side-by-side choices for eight copied voices on one 15-second excerpt. Positive errors are too high. The highest-note rule has mean absolute error 1.50 semitones, compared with 6.00 for median matching. These are within-author judgments, not population estimates.}

\label{fig:register}

\end{figure}

\section{Discussion and limitations}

\textbf{Access to open audio tools.} ARRANGE began with a composer’s ambition to hear a convincing voice over his own accompaniment. Suno served as a musical quality reference in the author’s exploration. ARRANGE pursues a route built around open audio tools and published agent skills, aiming to lower the barriers to AI-assisted composition while keeping musical decisions accessible to the composer.

\textbf{Skills and creative control.} Available models still require coordination: notes, lyrics, pronunciation, and timing must pass between tools without losing the composer’s choices. Skills package this work as musical tasks with defined inputs, outputs, and approval points. The score and plan give the composer decisions to inspect; typed edits give the agent a bounded way to apply corrections. The benchmarks support improved local change correctness; the observed Fable gains suggest that frontier capability does not make task-specific instructions and typed edits redundant. Whether this structure reduces musicians’ workload or increases their experienced agency remains to be tested.

\textbf{Human judgment and calibration.} A voice sample alone did not guarantee convincing singing in these trials. Finding a suitable key could improve the same voice, and the author found some voices better suited to particular song styles. This latter observation needs testing across a broader repertoire. Evaluation should distinguish voice resemblance from singing quality and use shared examples to anchor high, medium, and low ratings. Consistent audition conditions also matter, as the key comparisons showed. The present single-listener evaluations do not establish agreement across listeners.

\textbf{Current limits.} The musical examples cover two pieces and one experienced composer. The model benchmarks use small, project-curated request sets with repeated draws; their statistical tests are exploratory. Notation skills and model setup remain access barriers, and the evaluated workflow includes commercial LLMs as well as open audio models. We have not measured setup effort, cost, cognitive load, or comparative revision time. Broader studies should examine these barriers and test whether agents consistently respect composer checkpoints. Making composition more accessible remains a design aim.

\textbf{Rights and release.} Permission to use or distribute musical recordings and sampled voices remains separate from access to the software. A consent record alone does not establish permission; non-consenting singers’ audio remains private. Release exclusions and their consequences for reproduction are described below.

\section{Conclusion}

ARRANGE organizes adding a singing voice to a recorded performance around readable musical intermediates and bounded transformations. The plan-revision results show that typed operations can improve exact-change correctness across the tested LLMs. The score-editing results also show remaining interpretation and execution failures, including differences between prompted operations and installed-skill use. The first-person musical evaluations motivate a workflow in which the agent proposes and renders, while the composer inspects, corrects, and approves. Broader evaluation with composers is needed to determine whether this structure improves agency and creative practice.

\section*{Data and code availability}

Code, stored evaluation records, and reproduction instructions are available at \href{https://github.com/htansetiawan/arrange}{github.com/htansetiawan/arrange}. The project page is \href{https://htansetiawan.github.io/arrange/}{htansetiawan.github.io/arrange}. Many rewritten model answers are truncated, and copyrighted fits, score screenshots, reference recordings, and some songs are withheld. Stored summaries can be recomputed, but some experiments cannot be fully re-scored or regenerated from the release.

The voice comparison contains 35 presentations; 30 are publicly released, including nine first-hearing pairs, other-system clips, and hidden repeats. Released-subset means are 95.4 for SoulX-Singer and 90.0 for Seed-VC; SoulX scored higher in seven of nine pairs (paired Wilcoxon p = 0.098). These differ from the full-study values in Section 5.4. The register experiment’s 185 presentations comprised 48 pilot clips, 60 shuffled blind clips, 14 blind rechecks, and 63 ordered comparisons; these are presentations, not independent listeners. The correction benchmark’s full-study and released-subset results are distinguished in Section 5.2.

\section*{Declaration on generative AI}

An LLM agent is a component of the system described. Henry Tan conceived ARRANGE, designed the studies, made the musical and listening judgments, and wrote the original manuscript. Eidee Tan developed and inspected the generated scores and proofread the manuscript. Under the author’s direction, Claude assisted with implementation, experiment execution, statistical computation, and checks of text and references. ChatGPT assisted with copyediting and checking consistency against the available repository and cited sources. The human authors are responsible for verifying the final manuscript and approving it for submission.

\end{document}